\documentclass[twocolumn,english,pra,superscriptaddress]{revtex4-1}

\usepackage{color}
\usepackage{amssymb}
\usepackage{amsthm}
\usepackage{graphicx}
\usepackage{epsfig}
\usepackage{subfigure}
\usepackage{amsmath,amssymb,extpfeil,subeqnarray}
\usepackage{enumerate}
\usepackage[colorlinks,citecolor=blue,urlcolor=blue,anchorcolor=blue]{hyperref}
\usepackage{color}
\usepackage{bm}

\newcommand{\be}{\begin{equation}}
\newcommand{\ee}{\end{equation}}
\newcommand{\ba}{\begin{eqnarray}}
\newcommand{\ea}{\end{eqnarray}}
\newcommand{\ban}{\begin{eqnarray*}}
\newcommand{\ean}{\end{eqnarray*}}

\begin{document}

\title{\Large\textbf{Itinerant ferromagnetism entrenched by the anisotropy of spin-orbit coupling in a dipolar Fermi gas}}
\author{Xue-Jing Feng}
\affiliation{School of Physics, Henan Normal University, Xinxiang 453000, China}
\author{Jin-Xin Li}
\affiliation{School of Physics, Henan Normal University, Xinxiang 453000, China}
\author{Lu Qin}
\affiliation{School of Physics, Henan Normal University, Xinxiang 453000, China}
\author{Ying-Ying Zhang}
\affiliation{School of Physics, Henan Normal University, Xinxiang 453000, China}
\author{ShiQiang Xia}
\affiliation{School of Physics, Henan Normal University, Xinxiang 453000, China}
\author{Lu Zhou}
\affiliation{Department of Physics, School of Physics and Electronic Science, East China Normal University, Shanghai 200241, China}
\author{ChunJie Yang}
\email{yangchj@hotmail.com}
\affiliation{School of Physics, Henan Normal University, Xinxiang 453000, China}
\author{ZunLue Zhu}
\email{zl-zhu@htu.edu.cn}
\affiliation{School of Physics, Henan Normal University, Xinxiang 453000, China}
\author{Wu-Ming Liu}
\affiliation{Beijing National Laboratory for Condensed Matter Physics, Institute of Physics, Chinese Academy of Sciences, Beijing 100190, China}
\author{Xing-Dong Zhao}
\email{phyzhxd@gmail.com}
\affiliation{School of Physics, Henan Normal University, Xinxiang 453000, China}

\begin{abstract}
We investigate the itinerant ferromagnetism in a dipolar Fermi atomic system with the anisotropic spin-orbit coupling (SOC), which is traditionally explored with isotropic contact interaction.We first study the ferromagnetism transition boundaries and the properties of the ground states through the density and spin-flip distribution in momentum space, and we find that both the anisotropy and the magnitude of the SOC play an important role in this process. We propose a helpful scheme and a quantum control method which can be applied to conquering the difficulties of previous experimental observation of itinerant ferromagnetism. Our further study reveals that exotic Fermi surfaces and an abnormal phase region can exist in this system by controlling the anisotropy of SOC, which can provide constructive suggestions for the research and the application of a dipolar Fermi gas. Furthermore, we also calculate the ferromagnetism transition temperature and novel distributions in momentum space at finite temperature beyond the ground states from the perspective of experiment. \end{abstract}
\date{\today}
\maketitle

\section{Introduction}
Itinerant magnetism along with its spatially localized counterpart constitutes an indispensable part in modern condensed matter physics and also leads a long history of the intricate problem concerning magnetism \cite{wagner2013introduction,kubler2017theory,PhysRev.168.686}. If we omit the complicated interaction in advance, the well-acknowledged Pauli susceptibility, Landau diamagnetism and de Hass-van Alphen effect \cite{wagner2013introduction,kubler2017theory} will naturally unravel from solving the single-atom Hamiltonian via a statistically approach. When the cumbersome lattice potential is taken into account \cite{PhysRev.168.686}, it is still far better understood compared to the spontaneously ordered ferromagnetism, antimagnetism and ferrimagnetism, which are generally believed to be attributed to the intrinsic interactions. The crucial roles played by the complicated interactions have been remarkably revealed from early studies of electric gas \cite{1929Bemerkung,PhysRev.140.A1645} with Coulomb potential and of the Stoner model \cite{stoner1938collective,stoner1933xxx} in transition metals with an effective contact interaction. With the rapid development of quantum gas in recent years, the isotropic short-range contact interaction proves to be the basis of many theoretical portraits of cold atomic phenomena, including the itinerant ferromagnetism \cite{PhysRevLett.95.230403,PhysRevLett.105.030405,PhysRevA.85.043624,PhysRevA.93.063629,HE2014477,PhysRevLett.110.230401,zintchenko2016ferromagnetism,doi:10.7566/JPSJ.90.024004} with its experimental verification \cite{jo2009itinerant,RN28} that awaits the final truth. \par
In contrast to most works related to itinerant ferromagnetism dominated by the contact potential, more and more people put focus on the long-range dipole-dipole interaction \cite{PhysRevA.85.033615,PhysRevB.87.184424,2017JPhB...50a5302S,PhysRevA.98.023635,hu2022collisional,guo2022driven}.  Most strikingly, the anisotropic and long-range dipole-dipole interaction can induce many novel quantum phases such as the supersolidity \cite{PhysRevB.89.174511}, charge and spin density waves \cite{PhysRevB.91.224504,PhysRevA.87.043604} in recently achieved polar molecules $^{40}$K$^{87}$Rb \cite{K2008A,Bo2013Observation,PhysRevLett.108.080405,K2010Dipolar}, $^{23}$Na$^{40}$K \cite{PhysRevLett.109.085301} and magnetic dipolar $^{161}$Dy \cite{PhysRevLett.108.215301,PhysRevX.6.031022}. Furthermore, in dipolar Fermi systems, the itinerant ferromagnetism can be enhanced even without the traditionally necessary contact interaction \cite{20301,PhysRevA.105.053312,PhysRevLett.103.205301} and the Fermi surfaces are also distorted from spherical ones \cite{PhysRevA.77.061603,PhysRevA.81.033601,PhysRevLett.103.205301}.\par

 When it is unveiled concerning the physics of the relativistic regime in Dirac's theory, the spin-orbit coupling (SOC) naturally appears in the Hamiltonian designating itself an imperative role in relativistic quantum mechanics. This relativistic effect is not solely privileged in extremely high-speed objects but also can be ingeniously stimulated in the cold atoms with sufficiently low velocities by achieving an effective SOC \cite{Victor2011,PhysRevLett.109.095301,science2016,PhysRevLett.117.235304,huang2016experimental,PhysRevX.6.031022,wang2021realization,zhang2016properties,lu2022dynamics,jiao2022bose,yang2022dynamics} which arises from a synthetic gauge field created by the interaction between atoms and the various laser fields. The artificial SOC was firstly achieved in Bose gas in the form of the one-dimensional Raman-induced SOC \cite{Victor2011}. Soon after, this creative method of generating SOC was extended to higher dimensions and other fermion systems \cite{PhysRevLett.109.095301,science2016,PhysRevLett.117.235304,huang2016experimental,wang2021realization}. This Raman-induced SOC is not only highly controllable in different dimensions but also in the anisotropic aspects by adjusting the intrinsic freedom of Raman lasers. Recently, a three-dimensional (3-D) anisotropic SOC was successfully produced in optical lattice \cite{wang2021realization}. Thus our theoretical model is intimately connected to the further experimental realization in the dipolar gas with SOC. \par

Thus far, rarely have there been enough explorations into this topic of itinerant ferromagnetism when both SOC and exotic interactions beyond contact potential are considered. In our previous works, ferromagnetism has been investigated in dipolar Fermi gas with Raman-induced SOC \cite{PhysRevA.105.053312} and simple  one-dimensional (1-D) SOC \cite{20301}. It is unveiled that the anisotropy of the dipolar interaction makes great contribution to the formation of ferromagnetism and distortions of Fermi surfaces. However, the anisotropy of SOC has not been investigated since most of the previous works were concentrating on the isotropic Rashba-type or Weyl-type SOC with good symmetries \cite{PhysRevB.94.115121,PhysRevB.96.235425,VIVASC2020166113}. We believe in that it will be also important to study the SOC with a broken rotational symmetry in the spin space to see what novel phenomena might occur. Furthermore, as we have introduced above, the experiments in SOC is very progressive through the manipulation of various types of Raman lasers. This potentially experimental realization gives us another important motivation in considering the model with an anisotropic SOC. Although the traditional Rashba SOC and the isotropic Weyl SOC do not promote the ferromagnetic transition, we demonstrate that an anisotropic SOC can enhance it. Besides, the other explorations of anisotropic SOC include the vortex chain and resonance in anisotropically spin-orbit coupled Bose Einstein condensation (BEC) \cite{PhysRevA.98.013617,PhysRevA.78.023616,PhysRevA.87.063630,liao2018anisotropic} and superfluid as well as the Fulde–Ferrell pairing in three-dimensional spin-orbit coupled Fermi gas \cite{PhysRevA.88.013612,liu2015three}. \par 
In this work, we have proposed a useful scheme to the experimental observation of itinerant ferromagnetism. This paper is organized as follows. In Sec. \uppercase\expandafter{\romannumeral2}, we derived our formalism in a Hartree-Fock theory. The results are displayed and discussed in Sec. \uppercase\expandafter{\romannumeral3}. Finally, a conclusion is given in Sec. \uppercase\expandafter{\romannumeral4}.

\begin{figure}[t!]
	\includegraphics[width=1\columnwidth]{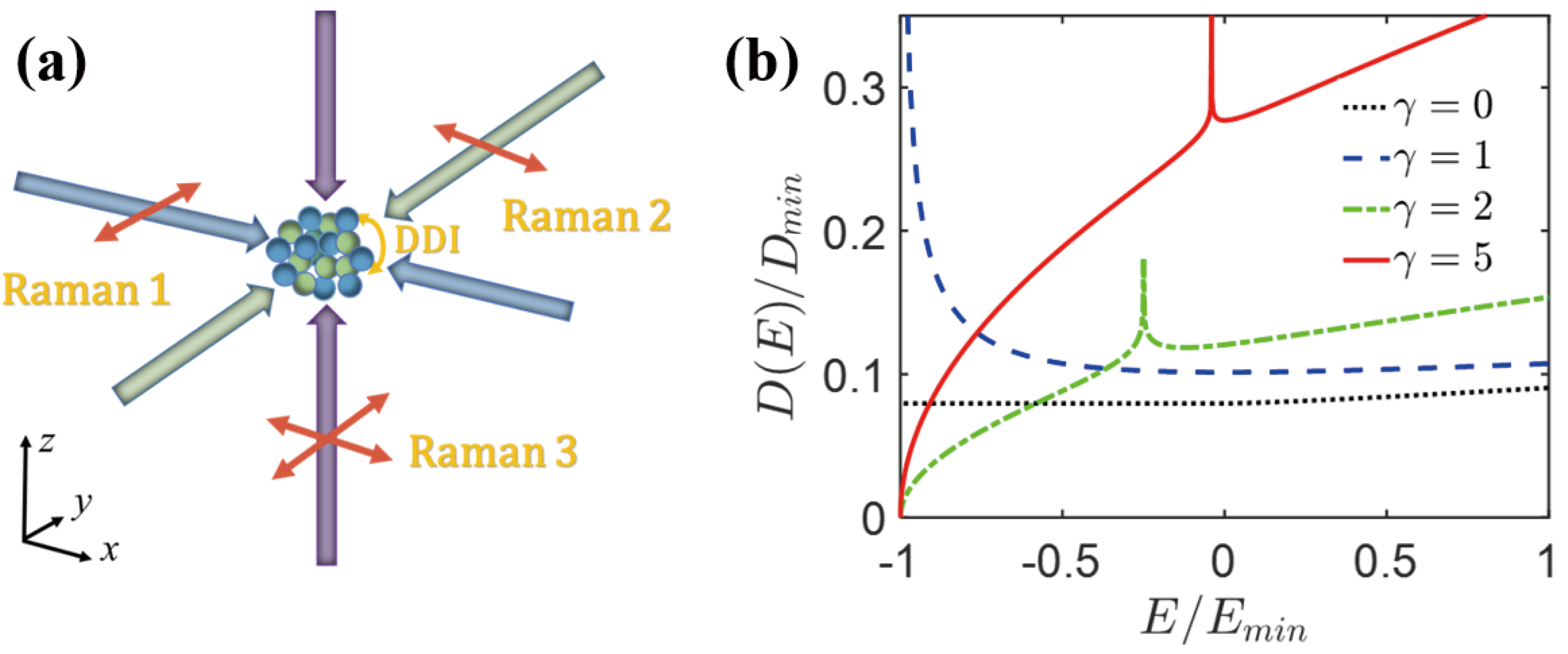}
	\caption{(Color online) (a) Schematic diagram of the experimental realization of the anisotropic 3-D spin-orbit coupling \cite{wang2021realization}. Three pairs of Raman lasers propagating in three directions are polarized in the directions displayed above. The $\emph{x}$-direction and the $\emph{z}$-direction Raman laser, along with the $\emph{y}$-direction and the $\emph{z}$-direction Raman laser, comprise a double-$\Lambda$ configuration leading to a 3-D SOC. The atoms interact with each other through a long-range anisotropic dipole-dipole interaction (DDI). (b) Density of states (DOS) of the lower-branch of single-atom excitation energy with different anisotropy parameters $\gamma$. The unit energy $\epsilon_{\lambda}=\hbar^{2}\lambda^{2}/(2m)$ with $\lambda=\alpha_{0}m/\hbar^{2}$. The unit of DOS $D_{min}=V\lambda^{3}/\epsilon_{\lambda}$. When $\gamma=0$ (Rashba SOC) or $\gamma=1$ (Weyl SOC), energy unit is $E_{min}=\epsilon_{\lambda}$. When  $\gamma=2$ and $\gamma=5$ which correspond to two anisotropic circumstances, the energy unit $E_{min}=4\epsilon_{\lambda}$ and $E_{min}=25\epsilon_{\lambda}$, respectively. The singularities of DOS with anisotropic SOC occur at $E=-\epsilon_{\lambda}.$ 
	}\label{fig1}
\end{figure}

\section{Model}

In our model, we consider an anisotropic 3-D spin-orbit coupling which is described by the Hamiltonian:
\begin{eqnarray}
H_{\rm SOC}&=\alpha_{0} (k_{x}\sigma_{x}+k_{y}\sigma_{y}+\gamma k_{z}\sigma_{z}),
\end{eqnarray}
in which $\sigma_{x}$, $\sigma_{y}$ and $\sigma_{z}$ are three Pauli matrices and $k_{x}$, $k_{y}$ and $k_{z}$ are three components of the wave vector, $\alpha_{0}$ represents the strength of SOC and the dimensionless parameter $\gamma$ represents the anisotropy of SOC. If we set $\gamma=0$, it returns to Rashba or Dresselhause SOC and if $\gamma=1$ we get an isotropic Weyl SOC. In this work we can continuously change the value of $\gamma$ to achieve different kinds of SOC. In fact,the parameters of  $\gamma$ is determined by wave vectors and Raman coupling of Raman lasers, which are highly tunable in experiments. So it’s a natural and practical way to achieve the anisotropic spin-orbit coupled Hamiltonian by performing the Raman coupling technique in cold atom systems. In a dipolar Fermi gas, the two-body interaction Hamiltonian $H_{I}$ includes both dipolar interaction and contact interaction which has the following form after second quantization:
\begin{small}
\begin{equation}
H_{I}=\frac{1}{2}\int d^{3}\mathbf{x}d^{3}\mathbf{x^{\prime}}\psi_{\alpha}^ {\dagger}(\mathbf{x})\psi_{\beta}^{\dagger}
(\mathbf{x^{\prime}})U({\mathbf{x},\mathbf{x^{\prime}}})_{\alpha\alpha^{\prime},\beta\beta^{\prime}} \psi_{\beta^{\prime}}(\mathbf{x^{\prime}})\psi_{\alpha^{\prime}}(\mathbf{x}),
\end{equation}
\end{small}
where $\psi_{\alpha}$ ($\psi_{\beta}$) and $\psi_{\alpha}^{\dagger}$ ($\psi_{\beta}^{\dagger}$) are fermion annihilation and creation operators for the $\alpha$ ($\beta$) component with $\alpha$ = $\uparrow$ ($\downarrow$) representing spin-up (spin-down) and
\begin{small}
\begin{equation}
U({\mathbf{x},\mathbf{x^{\prime}}})_{\alpha\alpha^{\prime},\beta\beta^{\prime}}=\frac{d^{2}}{r^{3}}\sigma_{\alpha \alpha^{\prime}}^{i}(\delta_{ij}-3\hat{\mathbf{r}}_{i}\hat{\mathbf{r}}_{j})\sigma_{\beta \beta^{\prime}}^{j}+g\delta_{\alpha\alpha^{\prime}}\delta_{\beta\beta^{\prime}}\delta(\mathbf{r}),
\end{equation}
\end{small}
where $\hat{\mathbf{r}}\equiv (\mathbf{x}-\mathbf{x^{\prime}})/\mid \mathbf{x}-\mathbf{x^{\prime}}\mid$, and $d$, $g$ are the dipole moment of the fermions and the coupling strength of the contact interaction. \par

Under the mean-field approximation, the total Hamiltonian in momentum space can be rewritten as:
\begin{align}
H=\sum_{\mathbf{k}}(a_{\mathbf{k},\uparrow}^{\dagger},a_{\mathbf{k},\downarrow}^{\dagger})\begin{pmatrix}
\epsilon_{1}(\mathbf{k})&\epsilon_{3}(\mathbf{k})\\
\epsilon_{3}^{*}(\mathbf{k})&\epsilon_{2}(\mathbf{k})
\end{pmatrix}
\begin{pmatrix}
a_{\mathbf{k},\uparrow}\\a_{\mathbf{k},\downarrow}
\end{pmatrix},
\end{align}
where $a_{\mathbf{k},\alpha}^{\dagger}$ and $a_{\mathbf{k},\alpha}$ are creation and annihilation operators in momentum space and
\begin{subequations}
\begin{align}
\epsilon_{1}(\mathbf{k})=&\frac{\hbar^{2}\mathbf{k}^{2}}{2m}+\alpha_{0} \gamma k_{z}+\frac{g}{V}\sum_{\mathbf{k}^{'}}n_{\mathbf{k}^{'},\downarrow}
\nonumber \\ 
&+\sum_{\mathbf{k^{'}}}\frac{4\pi d^{2}}{3V}\left[\left(3\cos^{2}\theta_{\mathbf{k-k^{'}}}-1\right) \left(n_{\mathbf{k^{'}},\downarrow}-n_{\mathbf{k^{'}},\uparrow}\right)\right],\\
\epsilon_{2}(\mathbf{k})=&\frac{\hbar^{2}\mathbf{k}^{2}}{2m}-\alpha_{0} \gamma k_{z}+\frac{g}{V}\sum_{\mathbf{k}^{'}}n_{\mathbf{k}^{'},\uparrow}\nonumber \\ &+\sum_{\mathbf{k^{'}}}\frac{4\pi d^{2}}{3V}\left[\left(3\cos^{2}\theta_{\mathbf{k-k^{'}}}-1\right)\left(n_{\mathbf{k^{'}},\uparrow}-n_{\mathbf{k^{'}},\downarrow}\right)\right],\\
\epsilon_{3}(\mathbf{k})=&\alpha_{0} k_{\rho}\exp(i\phi_{0})-\frac{g}{V}\sum_{\mathbf{k}^{'}}t_{\mathbf{k}^{'}}^{*}\nonumber \\
&+\sum_{\mathbf{k^{'}}}\frac{4\pi d^{2}}{3V}\left(3\cos^{2}\theta_{\mathbf{k-k^{'}}}-1\right)t_{\mathbf{k^{'}}}^{*},
\end{align}
\end{subequations}
in which $k_{\rho}=\sqrt{k_{x}^{2}+k_{y}^{2}}$ and $\phi_{0}={\rm arctan}(k_{y}/k_{x})$ and $V$ is the volume of this system. We denote $n_{\mathbf{k},\alpha}=\langle a_{\mathbf{k},\alpha}^{\dagger} a_{\mathbf{k},\alpha}\rangle$ as the particle density of spin-up and spin-down in momentum space and $t_{\mathbf{k}}=\langle  a_{\mathbf{k},\uparrow}^{\dagger}a_{\mathbf{k},\downarrow} \rangle$ as the spin-flip density.\par

The Hamiltonian above can be diagonalized by performing a unitary transformation of $(b_{\mathbf{k},\uparrow}^{\dagger},b_{\mathbf{k},\downarrow}^{\dagger})=$ $(a_{\mathbf{k},\uparrow}^{\dagger},a_{\mathbf{k},\downarrow}^{\dagger})S$, where the transformation matrix $S$ can be written as:
\begin{align}
S=\begin{pmatrix}
u(\mathbf{k}) \exp(i\phi_{0}) & -v(\mathbf{k})\exp(i\phi_{0})\\
v(\mathbf{k}) & u(\mathbf{k})
\end{pmatrix},
\end{align}
and $u(\mathbf{k})^{2}$, $v(\mathbf{k})^{2}$ respectively equals to $1/2\pm \left[\epsilon_{1}(\mathbf{k})-\epsilon_{2}(\mathbf{k}) \right]/2$ $\times\sqrt{[\epsilon_{1}(\mathbf{k})-\epsilon_{2}(\mathbf{k})]^{2}+4|\epsilon_{3}(\mathbf{k})|^{2}}$.
\par

Now the Hamiltonian can be diagonalized as
\begin{align}
H=\sum_{\mathbf{k}}(b_{\mathbf{k},\uparrow}^{\dagger},b_{\mathbf{k},\downarrow}^{\dagger})\begin{pmatrix}
\xi_{1}(\mathbf{k})&0\\
0&\xi_{2}(\mathbf{k})
\end{pmatrix}
\begin{pmatrix}
b_{\mathbf{k},\uparrow}\\b_{\mathbf{k},\downarrow}
\end{pmatrix},
\end{align}
where $b_{\mathbf{k}}^{\dagger}$ and $b_{\mathbf{k}}$ are creation and annihilation operator of the quasi-particles and
\begin{subequations}
\begin{align}
&\xi_{1}(\mathbf{k})=u(\mathbf{k})^2\epsilon_{1}(\mathbf{k})+v(\mathbf{k})^2\epsilon_{2}(\mathbf{k})+2u(\mathbf{k})v(\mathbf{k})|\epsilon_{3}(\mathbf{k})|,\\
&\xi_{2}(\mathbf{k})=v(\mathbf{k})^2\epsilon_{1}(\mathbf{k})+u(\mathbf{k})^2\epsilon_{2}(\mathbf{k})-2u(\mathbf{k})v(\mathbf{k})|\epsilon_{3}(\mathbf{k})|,
\end{align}
\end{subequations}
And the function of $n_{\mathbf{k},\uparrow}$, $n_{\mathbf{k},\downarrow}$ and $t_{\mathbf{k}}$ can be solved self-consistently with the condition of
\begin{subequations}
\begin{align}
&n_{\mathbf{k},\uparrow}=u(\mathbf{k})^2f\left(\xi_{1}(\mathbf{k})\right)+v(\mathbf{k})^2f(\xi_{2}(\mathbf{k})),\\
&n_{\mathbf{k},\downarrow}=v(\mathbf{k})^2f(\xi_{1}(\mathbf{k}))+u(\mathbf{k})^2f(\xi_{2}(\mathbf{k})),\\
&N=\sum_{\mathbf{k}}(n_{\mathbf{k},\uparrow}+n_{\mathbf{k},\downarrow}),\\
&M=\sum_{\mathbf{k}}(n_{\mathbf{k},\uparrow}-n_{\mathbf{k},\downarrow})/N,\\
&t_{\mathbf{k}}=u(\mathbf{k})v(\mathbf{k})\exp(-i\phi_{0})\left[f(\xi_{1}(\mathbf{k}))-f(\xi_{2}(\mathbf{k}))\right],\\
&f(x)=\frac{1}{\exp[(x-\mu)/k_{B}T]+1}.
\end{align}
\end{subequations}
In this work, a Hartree-Fock self-consistent method is performed to study the itinerant ferromagnetism in a dipolar Fermi gas with anisotropic SOC. For convenience, we can introduce a set of dimensionless parameters including dipolar interaction
 parameter $\lambda_{d}=n d^{2}/\epsilon_{F}$, SOC parameter $\lambda_{\rm soc}=\alpha_{0}k_{F}/\epsilon_{F}$, contact interaction parameter $\lambda_{s}=gn/\epsilon_{F}$, and temperature parameter $\lambda_{T}=k_{B}T/\epsilon_{F}$, where $\epsilon_{F}$, $k_{F}$, $k_{B}$ are Fermi energy, Fermi wave vector, and Boltzmann constant, respectively.

\begin{figure}[t!]
	\includegraphics[width=1\columnwidth]{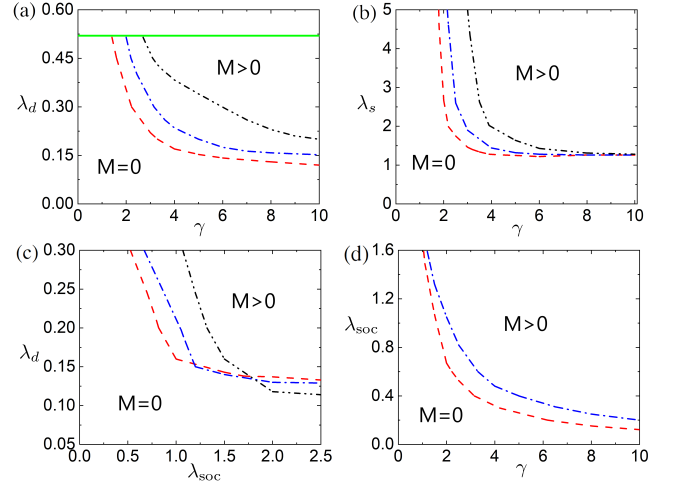}
	\caption{(Color online) (a) Zero-temperature ferromagnetism transition boundaries as functions of $\gamma$ and $\lambda_{d}$ with $\lambda_{s}=0$. (b) The same as (a), but as functions of $\gamma$ and $\lambda_{s}$ with $\lambda_{d}=0$. In panels (a) and (b), the red dashed line, blue dash-dotted line, and black dash-dot-dotted line are for $\lambda_{\rm soc}=0.6,\,0.4,\,0.2$, respectively. The green solid line in (a) is the unstable boundary above which the dipolar system undergoes a dynamical instability \cite{PhysRevLett.103.205301}. (c) The same as (a), but as functions of $\lambda_{\rm soc}$ and $\lambda_{d}$. In panel (c), the red dashed line, blue dash-dotted line, and black dash-dot-dotted line are for $\gamma=2.5,\,2,\,1.5$, respectively. (d) Zero-temperature ferromagnetism transition boundaries as functions of $\gamma$ and $\lambda_{\rm soc}$. In panel (d), the red dashed line and blue dash-dotted line are for $\lambda_{d}=0.3,\,0.2$, respectively. Above the boundary curves, this system undergoes a ferromagntic transition with the magnetization $\rm M>0$ from a normal state with $\rm M=0$. 
	}\label{fig2}
\end{figure}

\section{Results}
First of all, we can analyze some properties of the single-atom Hamiltonian with an anisotropic SOC by calculating the density of states (DOS). Two branches of excitation energies are $\epsilon_{k}^{\pm}=\hbar^{2}k^{2}/2m\pm\alpha_{0}\sqrt{k_{x}^{2}+k_{y}^{2}+\gamma^{2}k_{z}^{2}}$. The density of states by definition has the form of $D(E)=\sum_{\mathbf{k}}\delta(E-\epsilon_{k}^{\pm})$. The DOS can be calculated analytically and the results of the lower-branch are shown in Fig.~\ref{fig1}(b). When the anisotropy parameter $\gamma=0$ or $\gamma=1$, which correspond to the Rashba-type or the Weyl-type SOC, the DOS behaves continuously. However, when $\gamma>1$ which corresponds to an anisotropic SOC, the DOS has a singularity that is displayed in Fig.~\ref{fig1}(b). Here we want to emphasize that this singularity in DOS could be the origin of nontrivial properties in the following results that is concerned with the anisotropy of SOC.\par
The traditional Rashba and Weyl SOC don't promote any ferromagnetism transition without interaction. To achieve a ferromagnetic state in such systems with Rashba-type or Weyl-type SOC, we can only enlarge the strength of contact interaction or dipolar interaction to an extravagant extent, which seems to be an ineffective approach since the enormous interaction might invalidate our approximation and new physics might occur.The main theme of our present work is to find an alternative effective approach which is enormously helpful to an experimental observation. \par
As is indicated from Fig.~\ref{fig2}, when the anisotropy parameter $\gamma$ of SOC increases, the critical interaction responsible for ferromagnetism transition decreases significantly, which can provide a novel way of achieving the ferromagnetism for the further experiments. What has to be pointed out is, the contribution of $\gamma$ to the ferromagnetism seems to attain a saturation when $\gamma$ becomes sufficiently large, which means that although the anisotropy of SOC favors a ferromagnetic state, this enhancement will be converged to some extent. In Fig.~\ref{fig2}(a), the phase boundaries are plotted as functions of $\lambda_{d}$ and $\gamma$ without contact interaction and these curves decline significantly to small values. This promotion of $\gamma$ to ferromagnetism transition in a dipolar system seems to be an interesting result. As is acknowledged, the required strength of dipolar interaction that can induce the ferromagnetic transition in a simple dipolar system without any other SOC or external fields is near to the dynamical instability, about 0.51 \cite{PhysRevLett.103.205301}. So this significant reduction of the critical interaction by tuning $\gamma$ can be efficiently applied to the experimental observation of ferromagnetism in a dipolar system.\par

 Although the dipolar interaction in the present experiments in magnetic dipolar systems \cite{PhysRevLett.108.215301,PhysRevX.6.031022} is still not large enough, there are still promising approaches \cite{li2021tuning} to tune the effective dipolar interaction in some polar molecules. While in systems with contact interaction only, as is displayed in Fig.~\ref{fig2}(b), the results are not inspiring enough because the critical interaction are reduced to the saturation value near to that of the mean-field calculation of a system possessing a simple contact interaction. Here we have observed two kinds of interactions which are dipolar and contact interactions. Considering that the systems with contact interactions have been explored extensively in the previous works, in what follows we can reasonably ignore the contact interaction with $\lambda_{s}=0$ and reserve the dipolar interaction which will incubate interesting shapes of Fermi surfaces.\par
\begin{figure}
	\includegraphics[width=1\columnwidth]{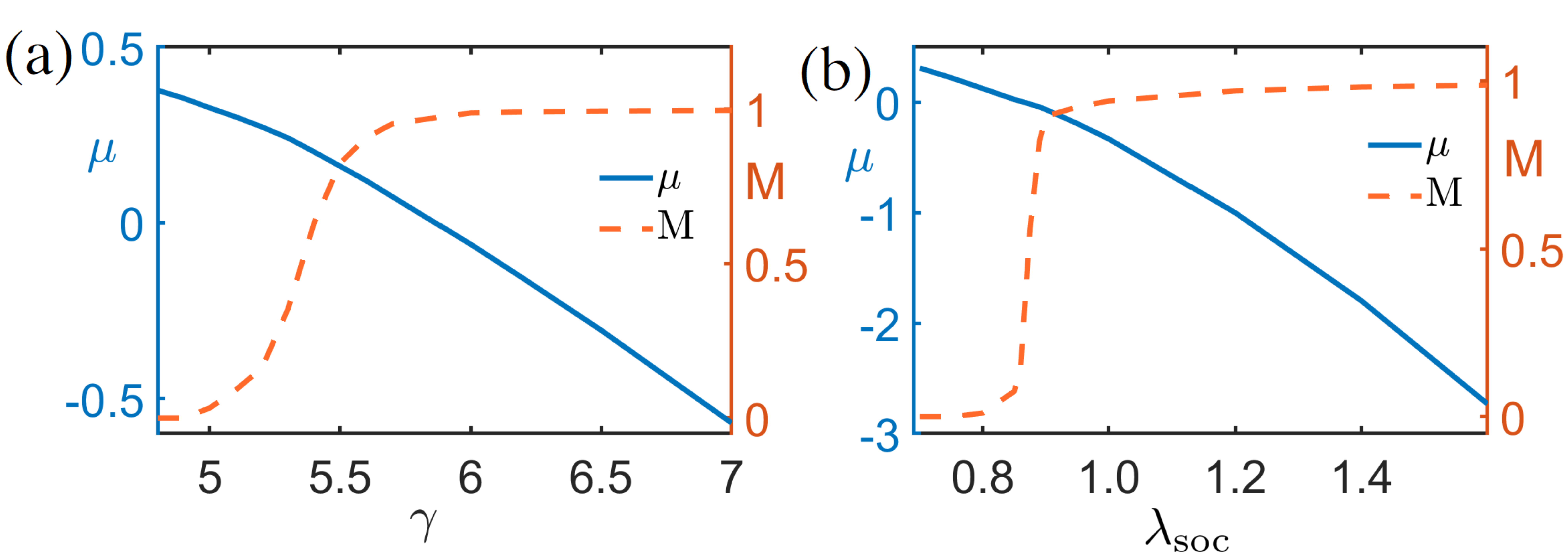}
	\caption{(Color online) Magnetization (M) and chemical potential ($\mu$) as functions of $\gamma$ with $\lambda_{d}=0.2$ and $\lambda_{\rm soc}=0.4$ (a); as functions of $\lambda_{\rm soc}$ with $\lambda_{d}=0.2$ and $\gamma=2$ (b). Both figures indicate a transition from a normal state (M=0) to a ferromagnetic state ($\rm M>0$). }\label{fig3}
\end{figure}
\begin{figure}[t]
	\includegraphics[width=1\columnwidth]{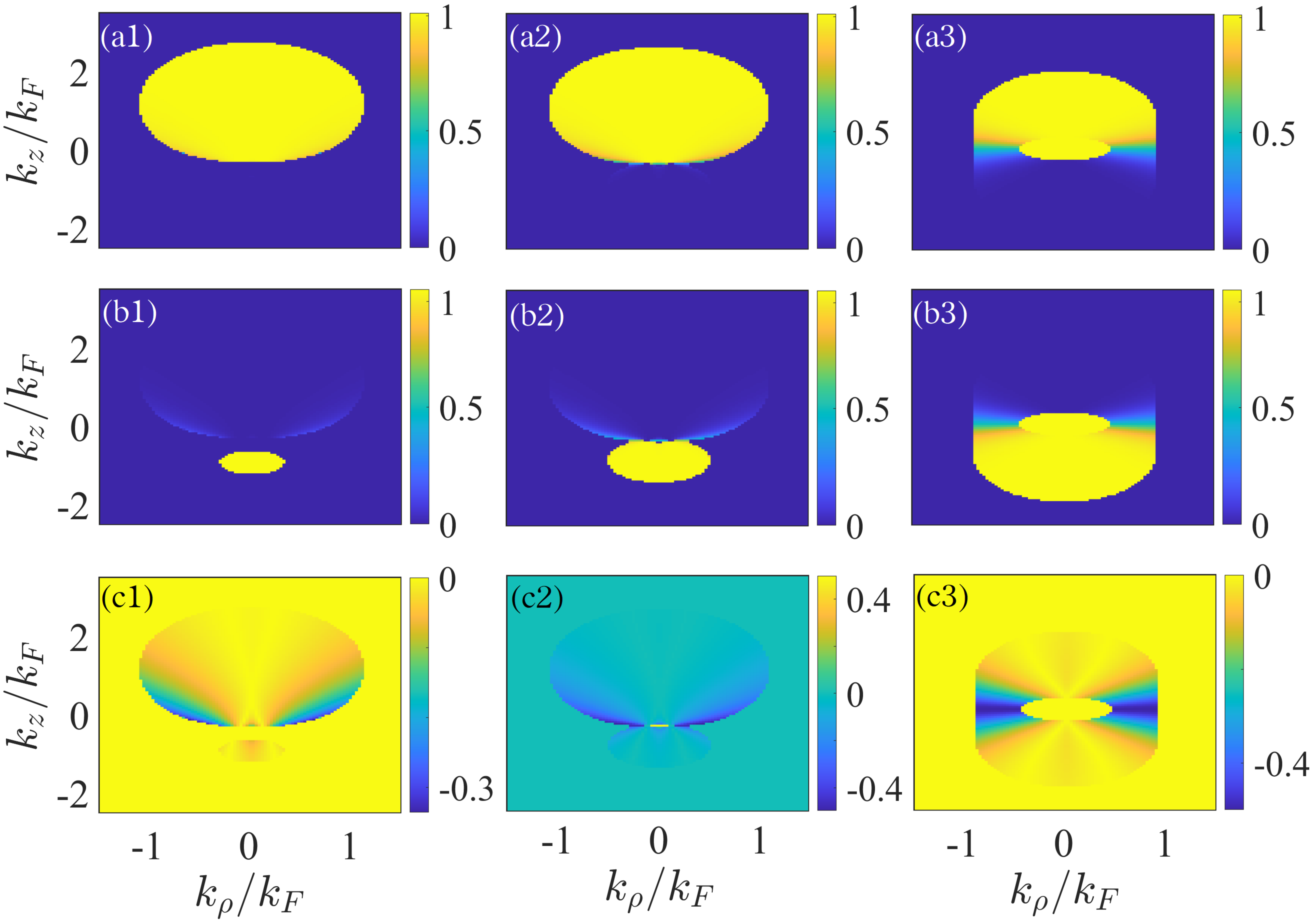}
	\caption{(Color online) Zero-temperature density distribution $n_{\mathbf{k},\uparrow}$ [panels (a1-a3)], $n_{\mathbf{k},\downarrow}$ [panels (b1-b3)], and spin-flip distribution $|t_{\mathbf{k}}|$ [panels (c1-c3)] with $\lambda_{\rm soc}$=0.4, $\lambda_{d}$=0.2. Panels (a1)-(c1) are for $\gamma$=5.7; panels (a2)-(c2) are for $\gamma$=5.5; panels (a3)-(c3) are for $\gamma$=4.9. These figures from the left column to the right column show a transition from a ferromagnetic state to a normal state.}\label{fig4}
\end{figure}
\begin{figure*}
	\includegraphics[width=2\columnwidth]{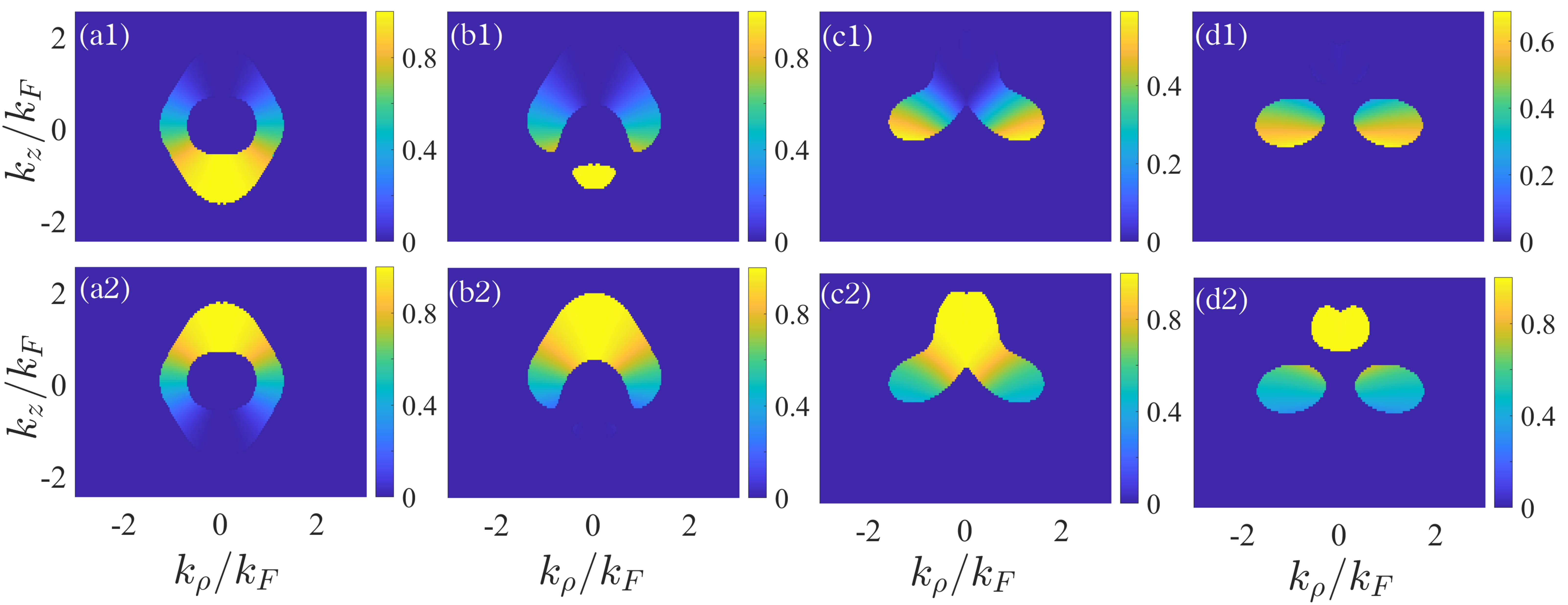}
	\caption{(Color online) Density distributions of spin-up [panels (a1-d1)] and spin-down component [panels (a2-d2)] in momentum space. Panels (a1-a2) are for $\lambda_{d}=0.1$; panels (b1-b2) are for $\lambda_{d}=0.13$; panels (c1-c2) are for $\lambda_{d}=0.4$; panels (d1-d2) are for $\lambda_{d}=0.5$, and all for $\lambda_{\rm soc}=1.9$, $\gamma=1.2$.}\label{fig5}
\end{figure*}
It can be concluded that the parallel SOC predominantly contributes to the polarization while the transverse component eliminates it, thus the anisotropic SOC will have a great impact on the phase transition. The enhancement of the anisotropy of SOC can be understood in another perspective. The SOC is essentially a momentum-dependent magnetic field thus the anisotropy indicates a favorable direction of a particular magnetic field which naturally results in the spin polarization. \par

\begin{figure}
	\includegraphics[width=1\columnwidth]{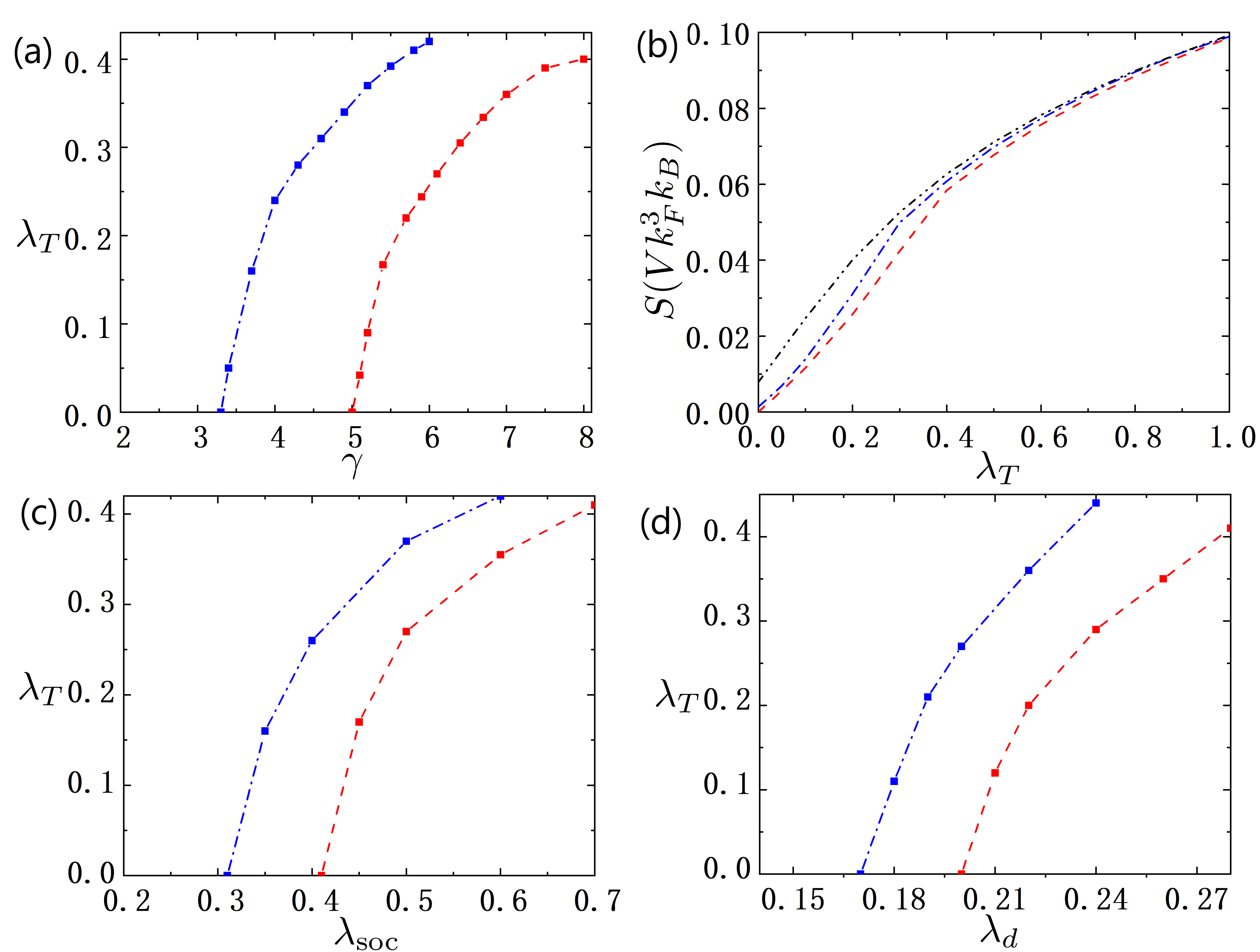}
	\caption{(Color online) (a) Ferromagnetic transition temperature as functions of $\gamma$ with $\lambda_{d}=0.2$. The red dashed line and  blue dash-dotted line are for $\lambda_{\rm soc}=0.4,\,0.6$, respectively. (b) Entropy as functions of temperature $\lambda_{T}$ with $\lambda_{d}$=0.20, $\lambda_{\rm soc}$=0.4. The red dashed line, blue dash-dotted line, and black dash-dot-dotted line are for $\gamma=8,\,6.5,\,5$, respectively. (c) Ferromagnetic transition temperature as functions of $\lambda_{\rm soc}$ with $\lambda_{d}=0.2$. The red dashed line and  blue dash-dotted line are for $\gamma=5,\,6$, respectively. (d) Ferromagnetic transition temperature as functions of $\lambda_{d}$ with $\gamma=5$. The red dashed line and  blue dash-dotted line are for $\lambda_{\rm soc}=0.4,\,0.5$, respectively.  }\label{fig6}
\end{figure}
\begin{figure*}
	\includegraphics[width=2\columnwidth]{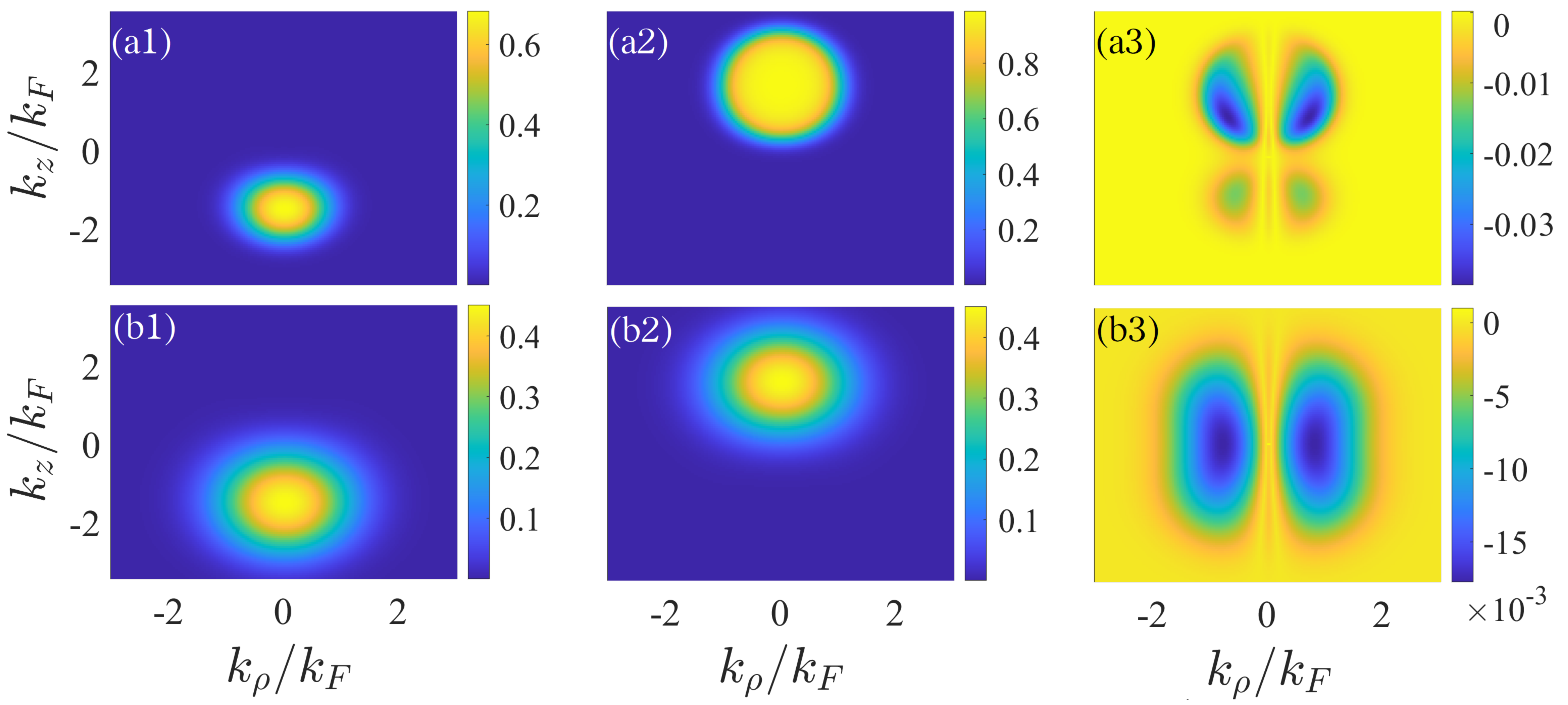}
	\caption{(Color online) Density distributions of spin-up [panels (a1-b1)], spin-down [panels (a2-b2)] and spin-flip distribution [panels (a3-b3)]. Panels (a1-a3) are for $\lambda_{T}=0.3$ and panels (b1-b3) are for $\lambda_{T}=1$. All panels are for $\gamma=8$, $\lambda_{\rm soc}=0.4$ and $\lambda_{d}=0.2$.}\label{fig7}
\end{figure*}
 At a fixed $\gamma$, when the strength of SOC increases, the critical dipolar interaction for ferromagnetism transition also declines, which are depicted in Fig.~\ref{fig2}(c). It also means that the strength of SOC can promote the ferromagnetism to some extent. Unexpectedly, at large $\lambda_{\rm soc}$, the critical dipolar interaction decreases when $\gamma$ becomes smaller, which indicates that the anisotropy cannot propel the ferromagnetism in this circumstance. We can call this phase region an abnormal area in the northwest corner of Fig.~\ref{fig2}(d) in which the phase boundaries are plotted as functions of $\lambda_{\rm soc}$ and $\gamma$ at fixed dipolar interaction. \par

We also calculate the magnetization and chemical potential when $\gamma$ and $\lambda_{\rm soc}$ increase and a ferromagnetic transition occurs, the results of which are shown in Fig.~\ref{fig3}. As displayed in Fig.~\ref{fig3}, the magnetization meets a sharp increase when the anisotropy parameter $\gamma$ and SOC strength parameter $\lambda_{\rm soc}$ pass through the transition points. This behavior of magnetization shows explicitly that both the anisotropy and the strength of SOC promote the spontaneous polarization as we have revealed in the previous discussion. The detail of this phase transition can also be unveiled by investigating the variation of chemical potential. As we can see from Fig.~\ref{fig3}, both the magnetization and chemical potential vary continuously and so does the deviation of the chemical potential, which indicates a second-order phase transition. \par

In Fig.~\ref{fig4}, the density functions and spin-flip distribution are displayed when it comes across a phase transition. As we can see, the Fermi surfaces in a ferromagnetic state are two unequal distorted ellipsoids. When it evolves into a normal state, Fermi surfaces are symmetric bowl-like shapes with a tiny ball embellishing aside. What has to be pointed out is that this grotesque shapes of Fermi surfaces seems omnipresent throughout our calculations. We can attribute this weirdness in Fermi surfaces to the exotic dipolar interaction since no such results have been reported in systems with isotropic contact interaction. This density distributions provide a powerful tool to investigate properties of our system, which we will see in the following content.\par
As we have pointed out, there exists an abnormal phase region in Fig.~\ref{fig2}(d) with large $\lambda_{\rm soc}$. It will be helpful to explore this region through calculating density distributions and the results are displayed in Fig.~\ref{fig5}.
At fixed $\gamma$ and $\lambda_{\rm soc}$, when we gradually enlarge the dipolar interaction parameter $\lambda_{d}$, we can see the variation of magnetization and Fermi surfaces. At small $\lambda_{d}$, this system is in a normal state and the Fermi surfaces are exhibited in the shapes of Figs.~\ref{fig5}(a1) and (a2). Here we have to emphasize that the Fermi surfaces have a rotational symmetry along the $z$-direction in momentum space thus the Fermi surfaces in the normal state at the beginning are topologically homeomorphic to spherical ball after completely rotating them around $z$-direction in $k$-space. It becomes totally unexpected when we add the dipolar parameter slightly when this system undergoes a phase transition from a normal state. The spin-up distribution in Fig.~\ref{fig5}(b1) consist of a ring and disconnected ball while the spin-down distribution in Fig.~\ref{fig5}(b2) is still a ball. They are topologically different in this circumstance. When we further augment the dipolar interaction, the Fermi surfaces evolve into arm-like shapes which are displayed in Fig.~\ref{fig5}(c1) and (c2). Finally, they revive into topologically different shapes in Fig.~\ref{fig5}(d1) and (d2).\par
The theoretical results above are concerning the ground states at zero temperature. However, in the experiment of a Fermi gas, the absolute temperature can not be possibly achieved and the system could be cooled down only to about 0.1$T_{F}$ to 1$T_{F}$. So the finite-temperature physics is of great importance to explore. As is indicated in Fig.~\ref{fig6}(a),
we calculate the ferromagnetism transition temperatures above which the normal states exist. We can see that the anisotropy of SOC greatly enlarges the critical temperature, which means that at a fixed experimental temperature tuning the anisotropy of SOC is still an effective way of observing the ferromagnetism transition. In Fig.~\ref{fig6}(c) and Fig.~\ref{fig6}(d), we calculated the transition temperatures as the functions of $\lambda_{\rm soc}$ and $\lambda_d$. We have found that transition temperatures increase with the dipolar interaction and the strength of SOC. At finite temperature, it is interesting to know how entropy behaves at different conditions. In Fig.~\ref{fig6}(b), we plot the entropy which takes the form of $S=-k_{B}\sum_{\mathbf{k}} \left[f(\mathbf{k})\ln f(\mathbf{k})
+(1-f(\mathbf{k}))\ln(1-f(\mathbf{k}))\right]$. The entropy increases with the temperature parameters, which accords with the general knowledge. We can also see from Fig.~\ref{fig6}(b) that the entropy at zero temperature is not always zero. Precisely, the entropy will attain zero at a ferromagnetic state and attain non-zero at the normal state. Thus we can distinguish the ferromagnetism transition by calculating the zero-temperature entropy. As far as we know, the zero-temperature entropy is connected to the degeneracy of ground states. So we can conclude that structures of a normal ground state and a ferromagnetic ground state are quite different. At large temperature parameters, we can also investigate how the particles are distributed in momentum space by calculating the order parameters and the results are displayed in Fig.~\ref{fig7}. At the condition of $\lambda_{T}=1$, it is apparent that the density distributions of the spin-up and spin-down are fairly symmetric in a Gaussian shape which totally smears out the details of the exotic ground-state Fermi surfaces. Besides, thermal fluctuation also eliminates the ordered ferromagnetic instability leading to a symmetric normal state.\par

\section{Conclusion}
 For most of the dipolar systems such as the magnetic $^{52}$Cr gas \cite{PhysRevLett.95.150406}, $^{167}$Er gas \cite{aikawa2014observation} and $^{161}$Dy gas \cite{PhysRevLett.108.215301}, the magnetic moments of them are 6$\mu_{B}$, 7$\mu_{B}$, and 10$\mu_{B}$, respectively. The dipolar molecules have much larger electric dipole moments, for instance, 0.566 Debye in polar molecules of KRb. Recently, a Raman spin-orbit coupled dipolar system was achieved in $^{161}$Dy atoms \cite{PhysRevX.6.031022}. Thus the manipulation of an anisotropic SOC will also be achievable following the previous approaches. The prominent results of this work are the predictions of the ferromagnetism transition under different parameters. To detect a spin polarization experimentally, we can monitor the suppression of collision because collisions would be forbidden in a fully ferromagnetic state, which might be easily achieved experimentally. Besides, in one of the previous experiments, a probing of the spin-dipole dynamics might also be an adopted way to demonstrate the spin susceptibility \cite{RN28}. To observe the particle distributions in the momentum space, we can apply the common techniques of expansion method. We can conclude that our theoretical work is much helpful to the experiments in the near future concerning both Raman SOC and dipolar interaction. \par
 In summary, we have explored the itinerant ferromagnetism in a dipolar Fermi gas with anisotropic SOC. It is confirmed after a Hartree-Fock calculation that the strength of SOC and the anisotropy of SOC can greatly reduce the critical interaction,above which a ferromagnetic phase transition will be found. By investigating the density distributions of the fermions, an abnormal behavior is demonstrated with its eccentric Fermi surfaces. In the last part, we also calculated the phase diagrams and order parameters at finite temperature and as we expect, the thermal excitation disorganizes the ground-state order leading to the common thermal disturbance. Our results can be promisingly applied to the further experimental schemes concerning this topic.

\section{acknowledgment}
Thank Xi-Bo Zhang for helpful dicussions about the experimental realization. This work was supported by the National Key R$\&$D Program of China under grants No. 2021YFA1400900, 2021YFA0718300, 2021YFA1400243, NSFC under grants Nos.12074105;12104135;61835013.

\bibliography{FOP.bib}



\end{document}